**Deep decarbonization and the Supergrid – prospects for electricity transmission between Europe and China**

Lina Reichenberg, Fredrik Hedenus, Niclas Mattsson, Vilhelm Verendel

ABSTRACT

Long distance transmission within continents has been shown to be one of the most important variation management strategies in renewable energy systems, where allowing for transmission expansion will reduce system cost by around 20%. In this paper, we test whether the system cost further decreases when transmission is extended to intercontinental connections. We analyze a Eurasian interconnection between China, Mid-Asia and Europe, using a capacity expansion model with hourly time resolution. The model is constrained by an increasingly tighter global cap on $CO_2$ emissions in order to investigate the effect of different levels of reliance on variable sources. Our results show that a supergrid option decreases total system cost by a maximum of 5%, compared to continental grid integration. This maximum effect is achieved when (i) the generation is constrained to be made up almost entirely by renewables, (ii) the land available for VRE farms is relatively limited and the demand is relatively high and (iii) the cost for solar PV and storage is high. The importance of these two factors is explained by that a super grid allows for harnessing of remote wind-, solar- and hydro resources as well as management of variations, both of which are consequential only in cases where dispatchable resources are limited or very costly. As for the importance of the cost for storage, it represents a competing variation management option, and when it has low cost, it substitutes part of the role of the supergrid, which is to manage variations through long-distance trade. The cost decrease from a Eurasian supergrid was found to be between 0% and 5%, compared to the cost in the case of continental-scale grids. We conclude that the benefits of a supergrid from a techno-economic perspective are in most cases negligible, or modest at best.

1. Introduction

The idea of a global super grid was launched in the 1970s by Buckminster Fuller [1]. In more recent years the rapid expansion of variable renewable power sources has spurred a renewed vision about intercontinental connections, see Battaglini et al. [2], Chatzivasileiadis et al. [3] and the EU Commission [4]. There are several potential reasons for a pan-continental grid, some of which are of a more political nature, such as international cooperation [5], or that it is claimed to be part of China's geo-political strategy [58]. Others argue that a global renewable grid would strengthen the global economy [8]. Finally, there are reasons pertaining to technical/physical properties: mainly the possibility of harnessing remote wind- and solar resources and to exploit spatial smoothing of wind- and solar output in high-VRE systems by transmission and trade [6][1]. The main proponent of a pan-continental grid is currently China, where Liu Zhenya, the president of State Grid, in 2018 said "From 2020 to 2030, the task will be to promote intracontinental interconnection with the interconnection of Asian, European and African grids being basically realised," [58].

---

[1] For a recent overview of possibilities and obstacles see [56].

In the scientific literature, the main focus has been on the geographic smoothing effect, for which spatial correlation of wind- and solar output is important. Although both solar and wind output are highly correlated over short distances, facilitating trade by building transmission over longer distances can smoothen variations in wind power supply to meet demand in different locations [7]. For a pan-continental grid, trade between several time zones may potentially add considerable value also for the day-night smoothing of solar power output [4, 9]. For renewable power systems at a continental scale, it has indeed been shown that building transmission to enable sharing of resources and trading of variations is a cost-effective strategy. This has been shown in model studies of Europe [10], South-East Asia [11, 12], South America [13] and North America [14, 15]. In these studies, large-scale transmission, rather than large-scale storage, is shown to be the most cost-effective solution for systems dominated by variable renewables. In cases with high shares of variable renewable electricity, the option of large-scale transmission typically reduces system cost by 10-30%, compared to the cost when countries are isolated [13, 16-18].

For an intercontinental grid there are, for similar reasons, suggestions and back-of-the-envelope calculations arguing for the cost-effectiveness of such a grid [3, 19, 57]. There are also a few model-based papers which investigate the interconnection of two or three continents: Brinkerink et al. [20] , Purvins et al. [21] and Blakers et al [22] investigate the cost-effectiveness of a Europe-US transmission cable in predominantly thermal power systems. The value of cables in such thermal systems is shown to pertain mostly to sharing the lower running costs of the US thermal power plant fleet. Dahl et al. [23] and Krutova et al. [6] investigate the interconnection of Europe-Asia-North Africa in highly renewable futures. Dahl et al. [23] assume that demand is covered entirely by wind and solar generation and that transmission extensions is the only variation management strategy. They find that, from a European perspective, interconnections to North Africa and Russia are cost-effective, while interconnections further east are not. Krutova et al. [6] focus on minimizing back-up generation and find that back-up energy may be greatly reduced by connecting continents, while back-up capacity is reduced only to a small extent. To our knowledge, the only study that compares the techno-economic benefits of a pan-continental grid to one with isolated continent-size regions is the recent paper by Breyer et al. [24]. They evaluated the effect of integrating nine "major regions"[2] of the world in a potential global grid and find that the cost benefit of doing so is a mere 2%.

The literature thus suggests that there are cases where inter-continental connections are beneficial or cost-effective. However, most studies suffer from at least one of two main methodological weaknesses: 1) they do not economically optimize investment and dispatch of a technology portfolio[3] and 2) they do not include other variation management strategies [25], such as energy storage, that could compete with transmission extensions to decrease cost.

To address these limitations in the previous literature on the subject, we use an optimization model with hourly time resolution, representation of storage and detailed supply curves for solar-, wind and hydropower generated by a GIS preprocessing step with explicit spatial representation. In contrast to Breyer et al. [24], we evaluate the pan-continental supergrid option under different assumptions related to technology costs; availability of nuclear power; land availability for solar- and wind farms (which reflects social acceptance) and uncertainty regarding the increase of electricity demand . In addition, we specifically target the question of the physical reality likely to drive a cost decrease due to a pan-continental supergrid: variation smoothing or resource sharing.

---

[2]Roughly the size of a small continent, such as Europe.
[3]In other words, they do not take the classical approach of Capacity Expansion Models (CEMs).

## 2. Method

We compare system cost for scenarios where there is an option to build transmission also between "superregions" (i.e. clusters of model regions, here China, Mid-Asia and Europe) and scenarios where transmission expansion is allowed only between countries. We do this using a Capacity Expansion Model, which is predominantly greenfield[4], and has overnight investment. The investment model, including the methodology to construct input data, is described below, as well as in a separate paper [26]. The scenarios which we compare are listed in Table 2.

**2.1 Investment model**

When modelling large geographic areas, there is a trade-off between technical, spatial and temporal detail on the one hand, and run time, on the other. Time resolution has been shown to be essential when modelling highly renewable power systems [27], and the development in continental-scale models is to move towards a full-year representation using time steps of one hour or three hours [11-13, 18, 27-30]. The model runs in this paper were therefore generated in a model version with hourly time resolution.

The model optimizes investment and dispatch of the electricity sector. Since the focus is to evaluate the cost-effectiveness of a future system with intercontinental grid connections, rather than the pathway to reach such a system, we employ overnight investment in a greenfield optimization approach. The exception is hydropower, where currently existing hydropower plants are assumed to be still in operation in 2050 Technologies are mainly represented by their investment costs, running costs and hourly capacity factors per region and resource class (for wind, solar and hydro power). Costs and other key parameters for technologies available for investment are listed in Table 1. The model is constrained by a global cap on $CO_2$ emissions expressed in g $CO_2$ per kWh of annual electricity demand.

There are no unit commitment-, ramping- or minimum downtime/uptime constraints in the current version of the model, which is therefore purely linear. For this reason, our model may overestimate the flexibility of thermal technologies. However, Cebulla et al. [31] observe that the effect of thermal constraints is relatively small for systems with high share of renewables, which is the focus of our study. Our model is written in the Julia programming language using the JuMP optimization package and is available online at Github [51]. For a more detailed description of the model, see reference [26].

**2.1 Regions, transmission- and technology assumptions**

In the study we use three "superregions": Europe, Central Asia and China. These are further divided into 6 to 8 subregions, see Figure 1. Each subregion is assumed to be a copperplate regarding internal transmission. Regions differ from each other in terms of availability of renewable resources as well as hourly demand and renewable output profiles. Transmission is only allowed between neighboring regions. Transmission costs are identical to those in [12], and are based on distances between population-weighted regional centers, and whether the connection is entirely on land or partially marine.

---

[4] Greenfield signifies that no existing power plants nor transmission lines are assumed to be remaining, and thus the cost estimate is as though the system would be built from scratch. The exception here is hydro power, where existing capacities are assumed.

*Table 1 The technology costs used as input parameters to the modeling.*

| Technology | Investment cost [€/kW] | Fixed cost [€/kW/year] | Variable cost [€/MWh] | Fuel cost [€/MWh fuel] | Efficiency |
|---|---|---|---|---|---|
| Gas GT | 500 | 10 | 1 | 22 | 0.4 |
| Gas CCGT | 800 | 16 | 1 | 22 | 0.6 |
| Coal | 1600 | 48 | 2 | 11 | 0.45 |
| Biogas GT | 500 | 10 | 1 | 37 | 0.4 |
| Biogas CCGT | 800 | 16 | 1 | 37 | 0.6 |
| Nuclear | 5000 | 150 | 3 | 3.2 | 0.4 |
| Wind power | 1200 | 43 | 0 | - | - |
| Off-shore wind | 2300 | 86 | 0 | - | - |
| Solar PV | 600 | 16 | 0 | - | - |
| Solar PV rooftop | 900 | 20 | 0 | - | - |
| Solar CSP | 6000 (incl. 12h thermal storage ) | 35 | 0 | - | - |
| HVDC converter pair | 180 | 1.8 | 0 | - | 0.986 |
| HVDC cables (land based) | 0.612 [€/kW/km] | 0.0075 [€/kW/year/km] | 0 | - | 0.016 (losses per 1000 km) |
| Hydro | (site specific variable cost) | - | - | - | - |
| Battery | 150 [€/kWh$_{storage}$] | 1.5 [€/kWh$_{storage}$/year] | 0 | - | 0.9 (round trip) |

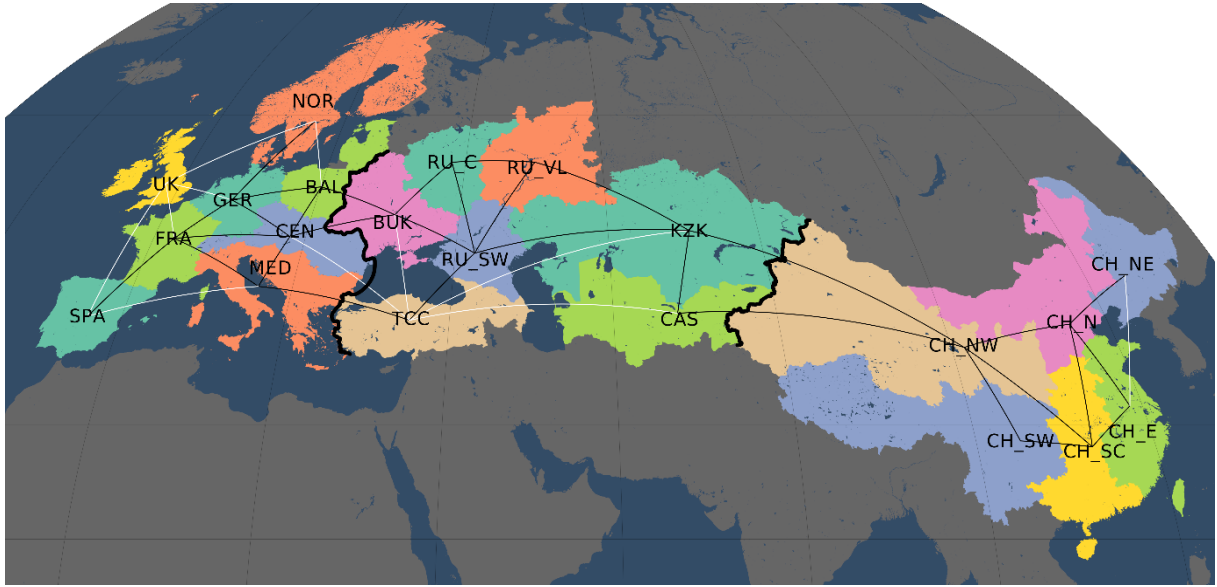

*Figure 1 Model regions. Subregions have different colors and are labeled with an abbreviation. Borders between superregions are marked with heavier black lines. Potential transmission connections are shown with thin black (land connection) or white lines (partially marine).*

## 2.2 Wind-, solar- and hydro data

Capturing regional differences in renewable supply curves and the correlation between temporal profiles of renewable output and electricity demand is of critical importance when modeling future near-zero carbon electricity systems. To estimate renewable profiles and supply potentials in a consistent manner, we perform a GIS preprocessing step based on ERA5 reanalysis data [32] combined with other public geospatial datasets for administrative borders [33], population [34], land cover [35], topography [36] and protected areas [37]. The general procedure is to use auxiliary datasets to remove pixels which cannot be used for large-scale wind- and solar plants, e.g. due to being in a forest or protected area, having too high population density (for onshore wind) or being situated in too deep water (for offshore wind). The remaining pixels are divided into five resource classes based on annual capacity factors, and finally we assume a certain fraction of each pixel in the remaining area is available for renewable power. Our baseline assumptions use 5 percent for solar PV and CSP, 8 percent for rooftop PV and onshore wind power, and 33 percent for offshore wind power. These numbers are congruent with a separate study made on current maximum land used for wind power, see Figure S11 in the Supplementary material. Due to the importance of these area assumptions and their inherent uncertainty we also perform an analysis in which the land areas where renewables can be deployed are doubled (see the scenarios labeled 'Hland' in Table 2). The areas are converted to potentials (in GW) for each region and resource class using typical power densities for solar and wind farms (Solar PV 45 MW/km$^2$, CSP 35 MW/km$^2$, onshore wind 5 MW/km$^2$ and offshore wind 8 MW/km$^2$). These GIS preprocessing assumptions and calculations are documented more extensively in [26].

Rooftop solar PV is estimated similarly to solar PV plants, except that it is only allowed in areas with high population density. To get a rough handle on potentials for remote solar and wind plants which would require additional transmission investments, we produce a proxy dataset for global grid access based on gridded population and purchase-power adjusted GDP for the SSP2 scenario in 2050 [38]. We assume that any 1x1 km grid cell with a local GDP exceeding 100.000 USD (2010) per year has direct grid access. This proxy is used to divide our five resource classes for solar PV and onshore wind power into two categories corresponding to local (within 150 km of the grid proxy) or remote (more than 150 km).

A visual representation of the classification of available land for wind and solar in three different subregions can be found in the Supplementary material (Figures S5 and S6).

Finally, we calculate the average capacity factor for each region, resource class and hour. Thus, an investment of, e.g. 1 GW of class 5 onshore wind power in a region will have an hourly output profile corresponding to the average profile of all class 5 wind power pixels in that region. The implicit assumption is that solar and wind plants with similar annual output are distributed evenly throughout each model region. For wind power, ERA5 hourly wind speeds are scaled by a factor that includes per-pixel annual average wind speeds from Global Wind Atlas version 3.0 [39] relative to long-term (1979-2019) per-pixel average wind speeds from ERA5. In this way we obtain an hourly time series of wind speeds that captures geographical variations in wind output caused by local differences in topography and land cover at a spatial resolution of 1 km (compared to 31 km for ERA5), while additionally preserving annual variations of wind abundance (which allows us to choose a data year that is more or less windy than average for the regions under study). Instantaneous wind speeds are converted into capacity factors using the output profile of the 3 MW Vestas V112 wind turbine, including wake losses and Gaussian smoothing to account for wind variations within a park. For concentrated solar power (CSP), hourly capacity factors for direct solar insolation are calculated assuming a solar collector multiple of 3.

For hydropower, we combine public databases on currently existing hydropower plants [40] and dams [41] with future potentials, costs, reservoir size and monthly inflow from [42]. All three datasets are geospatial with global coverage, which enables us to estimate existing capacity and future potential consistently throughout our model regions, including subregions of China. The scenarios reported in this paper, however, are restricted to use only existing hydropower capacity. The databases for existing hydropower cover approximately 80% of current plants globally, so we scale national totals to match existing capacity in [43]. Reservoir capacity in Europe is based on national statistics [52-54], and for regions outside Europe we (conservatively) set reservoir capacity to 6 weeks of peak hydropower generation. For regions where storage and run-of-river plants could not be distinguished due to lack of data, we limit hydropower flexibility by an aggregate minimum flow constraint (at least 40% of hourly water inflow must be used for electricity generation in each region).

**2.3 Demand data**

To comply with the aim of this study, it is necessary to use time-resolved demand data for all regions under study. Historic demand data with at least hourly resolution is required to model systems with high penetration of variable renewables with sufficient accuracy, and ideally the demand should use the same base year as input data for renewable supply to capture covariation due to e.g. correlation with hourly temperature variations. Additionally, development aspects need to be taken into account to represent our approximate time frame of 2050. However, hourly historic electricity demand data is not universally available, and therefore we chose to generate synthetic hourly electricity demand for our 21 model regions based on a machine learning approach. We train a gradient boosting regression model [44] to fit historic electricity demand in 44 countries for the year 2015 [45]. Regression variables include calendar effects (e.g. hour of day and weekday/weekend indicators), temperature variables (e.g. hourly temperature series in the most populated areas of each model region, or monthly averages and annual extremes as seasonality indicators) and economic indicators (e.g. local GDP per capita or annual electricity demand per capita). Cross-validation of the model predicts quantitatively and qualitatively similar time series for most countries on hourly and seasonal time scales, lending credit to the view that we have a model with generalization across larger geographical regions. The resulting hourly demand profile was scaled to match estimates of annual country-level annual electricity generation in 2050. These in turn were produced by extrapolating annual national demand in 2016 [59] using regional demand growth in the SSP2-26 scenario [38].

For more details of the implementation or evaluation, see Mattsson et al [26].

## 2.5 Scenarios and sensitivity analysis

We run a total of 12 scenarios, see Table 2. Four of these are our base scenarios, shaded in Table 2. The differences between these four base scenarios are: (i) possible connections between all regions ('Super', or 'S' in the scenario name) versus connections only being possible *within* superregions ('Continental', or 'C' in the scenario name) and (ii) a default more restrictive land available for wind and solar farms (no mention of land in the scenario name) versus more generous, high, land availability for wind and solar farms ('Hland' in the scenario name). In the four base scenarios we disallow nuclear power and CSP, but these are enabled in alternative scenarios (see sensitivity analyses below).

*Table 2 The characteristics of the different scenarios. For sensitivity scenarios, see below.*

| Scenario | Connections available | Land availability | Nuclear available? | CSP available? |
| --- | --- | --- | --- | --- |
| S | Super | Default | No | No |
| C | Continental | Default | No | No |
| S-Hland | Super | High | No | No |
| C-Hland | Continental | High | No | No |
| S-CSP | Super | Default | No | Yes |
| C-CSP | Continental | Default | No | Yes |
| R-CSP | None | Default | No | Yes |
| S-Nuc | Super | Default | Yes | No |
| C-Nuc | Continental | Default | Yes | No |
| R-Nuc | None | Default | Yes | No |
| R | None | Default | No | No |
| R-Hland | None | High | No | No |

The land availability assumptions for wind- and solar farms are:

- *'Default'*: 8%/5% (onshore wind/solar PV plants) of remaining[5] land is available for wind/solar exploitation

- *'High'*: 16%/8% (onshore wind/solar PV plants) of remaining land is available for wind/solar exploitation

In order to investigate the effect of the supergrid option under several technological futures, we run the base scenarios (rows 1-4, shaded in Table 2) for the lowest $CO_2$ cap of 1 g $CO_2$/kWh with different cost combinations of storage- and solar PV. For each of solar PV and storage, we assume three cost levels, for a total of nine cost combinations. The solar PV investment cost is set to 300 (*'low'*), 600 (*'mid'*), and 900 (*'high'*) €/kW and battery investment cost is set to 75 (*'low'*), 150 (*'mid'*), and 225 (*'high'*) €/kWh.

The reason why storage and PV are singled out among all input parameters for this treatment is twofold: First, these are the future costs to which some of the largest uncertainty has been attributed [46, 47]. Secondly, varying these two parameters allows us to investigate whether the benefit of spatial smoothing (through transmission) is nullified by the possibility of temporal smoothing (storage). One of the novelties of this paper is that significant spatial smoothing may be achieved also for solar, since our geographic scope spans eight time zones, while previous studies using capacity expansion models have spanned a maximum of four time zones.

In **Sensitivity analysis 1**, we investigate the effect of including nuclear and CSP, and check for the effect of isolating subregions. Scenarios with CSP are run in order to investigate the possible impact of this technology (rows 5 and 6 in Table 2). The reason why CSP is not an option in the base scenarios (rows 1-4 in Table 2) is that, in order to investigate the possible effect of varying the costs for solar and storage, we wanted to vary them independently. The CSP technology provides both generation and storage in the same technology, so its inclusion would cloud the effect of independent storage- and solar cost variation. As further comparison we also run two scenarios where investments can be made in nuclear power (one 'Continental' and one 'Super' connections available, rows 8 and 9. In addition, there are two scenarios (for default and high land availability) where subregions are isolated, so no transmission connections are available to invest in (rows 11 and 12). This scenario is labeled 'Regional' ('R' in the scenario name) and the purpose of it is to benchmark our results against those in literature investigating the benefit of connecting "country"-size regions to continental-size regions.

In **Sensitivity analysis 2**, we investigate the effect of increased demand. Here we base the construction of annual electricity demand (see Section 2.3 above) on the SSP1-45 scenario (25.2 PWh/year total over all three superregions), instead of the SSP2-26 scenario (19.7 PWh/year) in our base runs. Even though the storyline behind SSP1 is a more business-as-usual-like scenario for the future, such an increase in electricity demand may come about due to other factors, such as increased electrification of other sectors and demand for electro-fuels. Thus, sensitivity analysis 2 may be viewed as a general test regarding the effect of increased demand.

---

[5] See section 2.2. for the definition of "remaining land".

In **Sensitivity analysis 3**, we run the base scenarios using two additional weather years as input data, namely the years 1998 and 2003. These years were selected due to their higher (1998) and lower (2003) wind output in Europe and Mid-Asia, see Figure S10 in the Supplementary material[6].

3. Results
## 3.1 System cost and carbon caps

The system cost for the entire electricity system is investigated as a function of the carbon cap, with and without the supergrid option. As is apparent from Figure 2, the super grid option has nearly no effect on the system cost, compared to the cases where integration stops at the superregion level. Figure 2 shows the system cost for the electricity system for an increasingly tighter cap on $CO_2$ emissions, where the x-axis is the maximum $CO_2$ emissions allowed per kWh of electricity demand. Solar and storage costs are kept at their mid-level, see Table 1. The graph shows that the cost difference between the case where inter-continental transmission is possible (S) and the case where only connections within the continents are possible (C) is negligible. This is true for all carbon caps, even though a very small difference may be traced for tight carbon caps. In addition, the difference is small (<1%), both for the default case for land constraint (C versus S) and for the case with less severe constraints on land availability (C-Hland versus S-Hland).

As part of **Sensitivity analysis 1**, the model was run with halved investment costs for transmission (cables and converter stations). This does indeed increase the difference between the C and S cases. However, even with such unrealistically low costs for transmission, the cost difference amounts to a mere 5% (See Figure S1 in the Supplementary material).

**Sensitivity analysis 1** shows that, when nuclear power is an investment option, the benefit of a super grid is near zero for all $CO_2$ caps (see the Supplementary material, Figure S2). This is explained by that the benefit of trading is smaller when all subregions are given the option to invest in an unlimited, non-variable, $CO_2$ neutral technology. Notably, the regional scenario with default assumptions on land availability (R) displays a high cost for an emission cap of 10 g $CO_2$/kWh and is infeasible for emission caps below 10 g $CO_2$/kWh. (However, when nuclear is allowed, the R-scenarios are feasible, and only 5-7% more costly than the integrated scenarios S and C for targets below 20 g $CO_2$/kWh, see the Supplementary material, Figure S2.)

In **Sensitivity analysis 1**, we also run a regional scenario with high land availability assumptions (R-Hland), which is feasible even for a 1 g $CO_2$/kWh cap[7], but is considerably more costly than the integrated scenarios (S, S-Hland, C, C-Hland), see Figure 2. For the case of isolated regions with high land availability (R-Hland), the average system cost (at 1 g $CO_2$/kWh[8]) is 70.7 €/MWh, compared to 53.0 €/MWh for continent grid integration options (scenario C-Hland) and 52.5 €/MWh for the supergrid option (scenario S-Hland). In other words: the cost reduction due to integration of regions into continents is on average 25% (15% for Mid-Asia, 21% for Europe and 27% for China), while the cost reduction due to integration of continents into a supercontinent is almost none.

---

[6] However, as may be seen in Figure S10, 'high' and 'low' wind output is not trivial when the study area is as large as in this study: Europe and China actually appears to be even anti-correlated with respect to annual mean wind speed. Both 1998 and 2003 were average or even slightly below average years for wind output in China.
[7] Scenario R becomes infeasible for emission targets lower than 10 g $CO_2$/kWh.

The effect on system cost from the supergrid option is thus small for all the scenarios, including those with strict land constraints. However, the *absolute* cost level (S versus S-Hland and C versus C-Hland) is affected by more conservative estimates in land availability, with a cost difference of 9-10% for targets below 20 g $CO_2$/kWh (i.e. the difference between full and dashed blue and red lines in Figure 2). The results thus show that the amount of land available for wind- and solar farms is an important parameter for the cost of a renewable power system. However, the assumption on land availability has a negligible effect on the cost-effectiveness of a Eurasian supergrid.

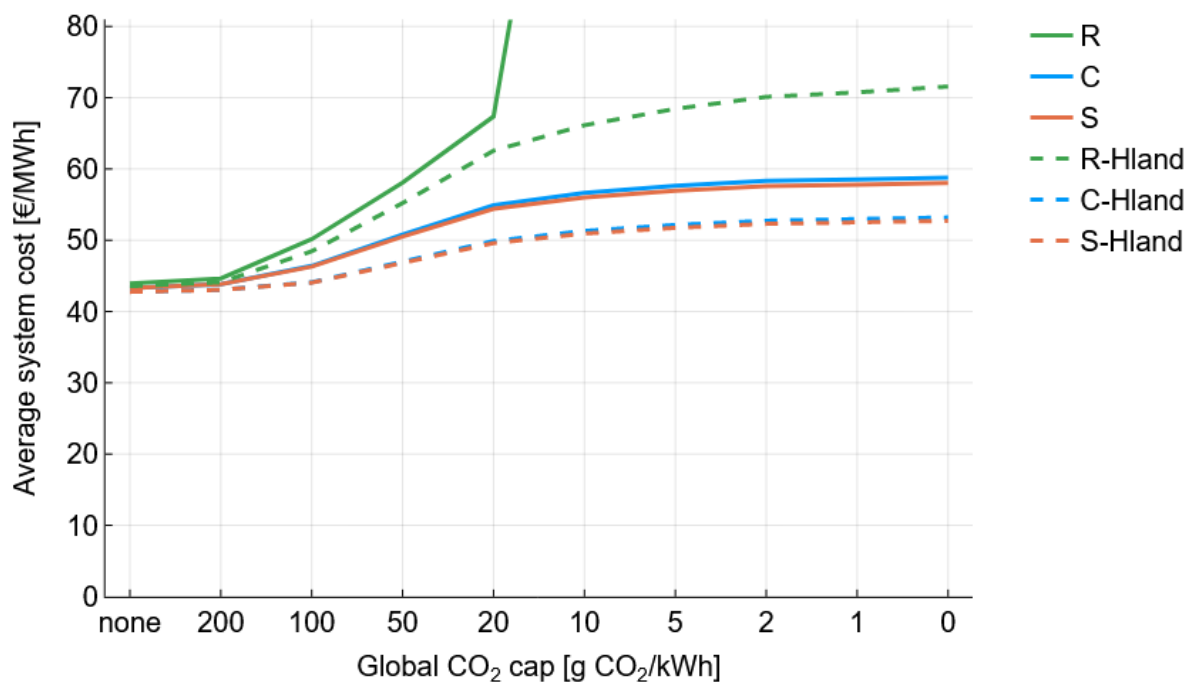

*Figure 2 The system cost of electricity for increasingly stringent carbon targets, in the cases of Regional ('R'), Continental ('C') and Superconnected ('S') transmission expansion availability. All cases were run for default land availability ('R', 'C', 'S') and high land availability ('R-Hland', 'C-Hland', 'S-Hland'). For further details regarding the scenarios, see Table 2.*

### 3.2 Electricity supply mix

In Figure 3, the annual electricity supply mixes in the four base scenarios (S, C, S-Hland, C-Hland in Table 2) are presented.

The figure reveals that an intercontinental supergrid has a negligible effect on the electricity supply mix (comparing scenarios S with C, and S-Hland with C-Hland, respectively). It is known from literature that wind is advanced by increased transmission capacity on a continental level [16, 24] and solar PV is advanced by decreasing cost for storage [48]. The tendency towards more wind power in the super-interconnected cases (S and S-Hland) is present also in our results, but the effect is very small, see Figure 3. The limited effect on capacity mix is explained by the fact that there is substantial amount of transmission *within* the superregions. The additional option for transmission investment between superregions thus has only a minor effect on the wind/solar mix.

The main difference in the supply mix is instead between cases with the default (S, C) and high (S-Hland, C-Hland) land availability. In the case of high land availability, more wind power is utilized at the expense of solar PV. This illustrates the fact that onshore wind power in many cases is the most cost-effective technology, but it is limited due to land constraints in the low land availability scenarios. The battery use, too, is higher in the default cases (S and C) with higher share of solar PV[9]. The reduced battery use for the high land availability cases (S-Hland and C-Hland) may be expected since batteries are primarily a short-term storage technology, with a natural synergy with the regular day/night cycles of solar PV.

From Figure 3, we can also see that the level of curtailment and loss in transmission and storage operations, which is the difference between demand (indicated by a black line) and the total primal generation (by wind, solar, hydro and bio), is similar between the cases and amounts to 10-11% of annual demand. (Electricity discharge from batteries should be disregarded for this comparison since that energy would otherwise be counted twice.)

---

[9] In the case where CSP is allowed (scenario C-CSP versus S-CSP and Figure S3 in the Supplementary material), the built-in storage function in the CSP plants reduces the utility of external storage (batteries) and gives CSP a comparative advantage to PV.

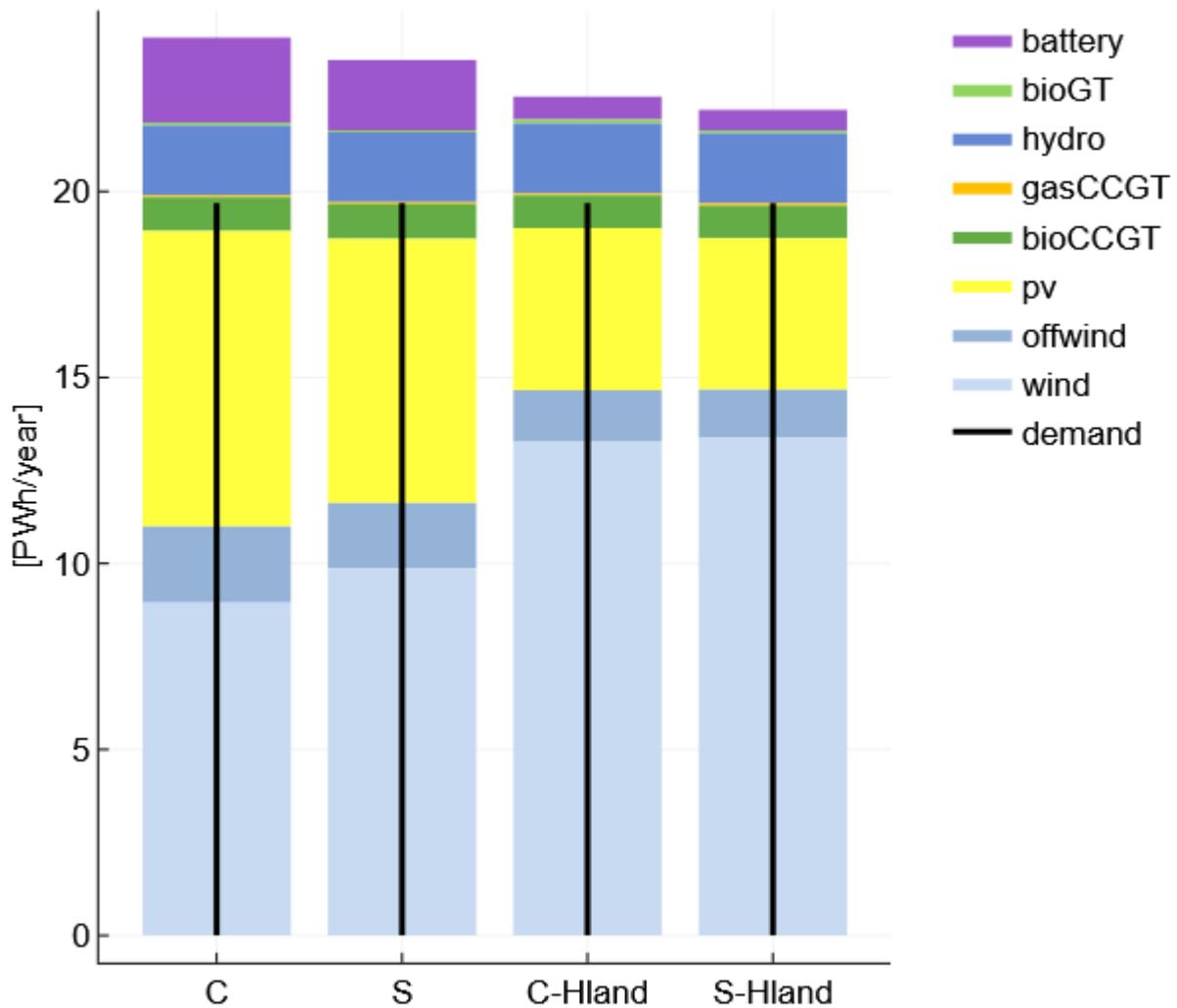

*Figure 3 The energy supply mix in the four different scenarios for a carbon constraint of 1 g $CO_2$/kWh. The black lines show the annual demand.*

### 3.3 Effect of solar and storage cost on the cost-effectiveness of a supergrid

Figure 4 shows how storage- and solar PV costs impact the cost reduction potential of a supergrid for a cap of 1 g $CO_2$/kWh. The dots in Figure 4 indicate the difference (in percent) between the costs for continental and supergrid scenarios, i.e. comparing the system cost for S with C and S-Hland with C-Hland.

The first observation is that the maximum effect of the supergrid option, compared to continental grids, is 4%. This maximum value of a supergrid occurs when the solar cost is high and the battery cost is high, i.e. when the main generation technology which is benefitted by long-distance transmission (wind) is comparatively less costly than solar, while at the same time the 'competing' variation management strategy (storage) is comparatively more costly than transmission.

Figure 4 shows that the cases with lower cost for storage always have, *ceteris paribus*, smaller cost reduction due to the super-interconnection option. This result is intuitive: the less costly the competing variation management strategy (storage), the less important the variation management strategy enabled by the supergrid option (intercontinental trade). The effect is further increased if solar PV also has low cost, as can be seen by following the vertical axes in Figure 4. As has been shown in previous studies, low-cost storage increases the share of solar PV, while not having the same effect on wind [17, 48]. Our results, which show that lower cost for solar PV decreases the effect of the supergrid option, are thus in line with previous results in literature based on continent-wide model studies.

The cases where the higher availability of land was assumed (Figure 4, right) show similar tendencies regarding the effect of decreasing solar PV- and storage cost. However, overall, the cost gain from the supergrid option is smaller, where the maximum effect is 1%. This is because a larger resource base means that there is less gain to be had from importing from afar, which is one of the mechanisms by which increased availability of transmission effects the system cost.

**Sensitivity analysis 2**, based on SSP1 where the total annual demand is approximately 25% higher than in the SSP2, shows that increased demand has a similar effect as decreasing the amount of available land. The maximum effect of the supergrid option was then 5%, see Figure S4 in the Supplementary material.

**Sensitivity analysis 3**, where two other weather years were used as input to the optimization, showed no significant difference regarding the effect on cost of the supergrid option, see Figures S5 and S6 in the Supplementary material.

In summary, even when all parameters are aligned to make the supergrid option cost effective[10], the impact of it is still only a mere 5%, compared to continental grids.

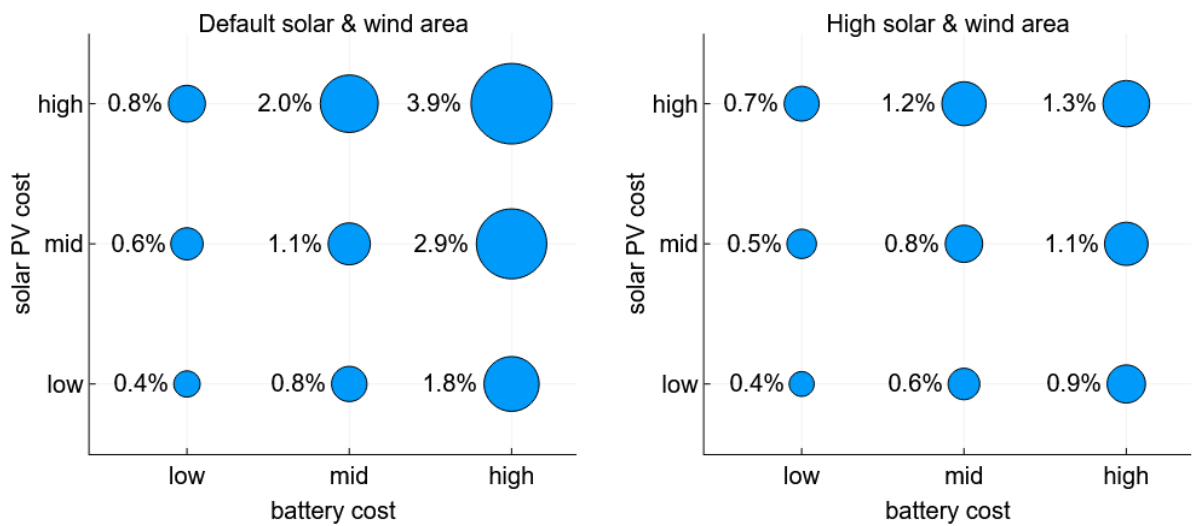

*Figure 4 The increase in total system cost for a system without interconnected superregions compared to with interconnected superregions, using varying assumptions for battery and solar PV costs. The left figure shows the difference between cases S and C, and the right side between cases S-Hland and C-Hland.*

---

[10] The maximum effect is achieved with the following settings: default land availability, high cost for solar PV and storage, a high demand (SSP1).

## 4. Discussion

This paper investigates the cost reduction from allowing for inter-continental grid connections between Europe and China, by using a Capacity Expansion Model with 21 regions in China, Mid-Asia and Europe. The results indicate that a Eurasian supergrid decreases system costs by 0-5%, where the highest cost reduction is achieved for cases with high cost assumptions for solar PV and battery storage, a high projected future demand (SSP1) and default land availability for solar and wind. We thus find that the effect from connecting continents has a much smaller effect on system cost than connecting individual countries in a continent-wide grid, which here was estimated to 15%, 21% and 27% for Mid-Asia, Europe and China, respectively. These results regarding cost reduction for country-to-continent integration is in line with those in literature, which are in the range of 15-30% [13, 16-18]. Both of these results, the decrease in system cost from allowing for pan-continental transmission and the decreasing marginal benefit from integration, are in line with the preliminary results by Breyer et al. [24]. They found that the possibility to connect nine large regions into one possibly global grid decreases system cost by 2%, and that the major benefit of electrical integration is achieved already with 23 global regions.

Thus, judging from our results, it seems that the marginal benefit of electrical integration significantly decreases when the integration region exceeds the size of a continent (Europe, Mid-Asia or China in our results). In order to explain this, we return to the two main motivations in literature for why a supergrid would potentially be cost-effective: tapping remote resources and managing variations [6].

Resources in this context refers to hydro-, wind- and solar resources, where hydro- and solar resources are approximately equally distributed among the superregions. Thus, for hydro- and solar power, there remains that the factor of the possible smoothing effect for solar by connecting several time zones, would contribute significantly to a cost reduction. This is, however, not confirmed by our results. On the contrary, less costly solar, and thus more solar in the mix, entails a smaller benefit for the Eurasian supergrid compared to continental grids, than the base case costs (Figure 4). This result shows that, for solar, storage is a more cost-effective variation management tool than is transmission and trade, even when transmission over several time zones is available as an option. Thus, the results from continent-size models regarding the synergy between storage and solar [55], which has to do with that the day-night cycle of solar requires smaller amounts of storage than do those for wind, still hold even if the span of time zones is considerably increased.

Compared to solar- and hydro resources, wind is more unevenly distributed between the superregions, especially in the default case regarding land availability, where China has less abundant wind resources in relation to its demand (Figure S7). Wind power is also known as being the beneficiary of variation smoothing from transmission in previous research on a continental scale [10, 13-15]. The results show a small increase in wind power capacity in the supergrid case (S) compared to the continental case (C) for the default land constraints, c.f. Figure 3. There is only a minor change in wind power capacity in the case with a higher land availability (S-Hland compared to C-Hland). Since the supergrid option has virtually no effect on the optimal share of wind power in the case less constrained regarding land availability, it is likely that the increase in wind power from the supergrid option has to do with resource tapping, rather than with variation management. Thus, it seems like the smoothing effect/variation management is achieved already by integrating *one* superregion, and hence the lack of effect on wind power from further electrical integration. This hypothesis is indeed concurrent with the knowledge of the spatial correlation of wind power output [8], and size and time scale of weather patterns.

The spatial heterogeneity regarding cost and availability of wind and solar power in our study, i.e. the fact that a country such as Kazakhstan may produce a large quantity of low-cost wind electricity while a region such as South-East China may not, is due to varying quality and quantity of the wind- and solar resource. However, in addition to the resource quality, the production cost depends on investment- and maintenance costs, as well as the interest rate, which are assumed to be equal between regions in the present paper. It may be reasonable to assume that these costs, too, vary by country and correlate with e.g. regional GDP. Thus, such variations may be an additional source of regional heterogeneity, and one that may drive additional value of a super grid. This interesting topic is, however, left for future work.

As a side result, we show that land availability, in contrast to the supergrid option, is consequential for the system cost of renewable power systems. The effect on system cost, both in the cases with multi-continent integration and with single-continent integration, from a doubling of land availability was a system cost reduction of 9-10%. Our assumptions on land were simple: we assumed all land not belonging to any of the "forbidden" categories (natural reserves, highly populated areas, lakes) to be exploitable to the same degree. Thus, even though our methodology does not allow for a detailed assessment of the effect of popular sentiment affecting the exploitation of wind- and solar resources, the results show that the land availability seems to have a relatively large impact on system cost. This effect is further exacerbated when continental integration was not allowed (the regional, or 'R' scenarios). These scenarios were infeasible for very low carbon caps, where the infeasibility stems from two densely populated regions in Eastern China. The issues surrounding availability of land are explored only to a limited extent in the modelling literature, so we anticipate more research going into a detailed assessment of this issue as well as its consequences for cost and feasibility of renewable power systems.

Our model is limited in technological scope, for example in that it does not include sector integration, which could (i) increase the volume of electricity consumption[11] and (ii) provide a market for "excess" wind and solar power to produce e.g. hydrogen. Regarding (i): Our **Sensitivity analysis 2** shows that increasing the demand has a similar effect as limiting the land available for wind and solar, thus creating a drive for connections to share resources. However, as the discussion above and Figure 2 show, scarcity of land for VRE deployment does not seem to significantly increase the cost benefit of a Eurasian supergrid. Thus, increased electricity demand is not likely to substantially alter the results regarding the cost-effectiveness of the supergrid. (ii) Sector integration, which, among other effects, leads to an increased demand for electro-fuels, has been seen to *decrease* the impact on system cost from transmission extensions on a continental scale (c.f. [10, 16]). This is explained by that sector coupling provides an additional variation management strategy, thus decreasing the importance of other such strategies, among them electricity trade. Hence, it is likely that a sector-coupled model, such as [28], would rather strengthen our main results regarding the minor cost benefit of connecting China and Europe.

## 5. Conclusions

This paper evaluates the cost benefit of a Eurasian supergrid, connecting Europe with China, via Mid-Asia in a renewable future. We show that a necessary condition for a Eurasian supergrid to have significant impact on system cost, is that the power system is close to carbon neutral and that dispatchable carbon-free generation such as nuclear power or CCS is only used to a very limited extent. We further show that the benefit even in a renewable system is still small, in the base case around 1%. Larger benefits, up to 5%, are obtained when solar PV- and storage costs are assumed to be high, and land availability for solar and wind low. Thus, even though our study ignores technical difficulties of large transmission extensions [49], or their possibly undesired geopolitical consequences [4, 50] of large transmission extensions, the supergrid option between Europe and China via Mid-Asia does not appear particularly attractive. This result stands in contrast to those who have argued that a pan-continental grid brings a large potential for cost reduction [1-3, 5, 6] However, none of these studies really evaluate the effect on cost from the super grid option compared to intra-continental transmission. Nevertheless, they argue optimistically regarding the benefits, including cost reduction, from a super grid. While our results are "negative", in the sense that inter-continental transmission expansion has relatively little effect on system cost, the benefit of substantially *more* integration than is currently the case is evident from our results concerning integration of subregions into continents. Thus, our results still confirm transmission as one of the important pieces of the puzzle to build an affordable renewable energy system.

---

[11] There are of course also other reasons that demand may increase beyond that of our prediction here.

SUPPLEMENTARY MATERIAL

Lina Reichenberg, Fredrik Hedenus, Niclas Mattsson, Vilhelm Verendel

The Supplementary material consists of additional model results. For a more detailed description of the model and input data, please refer to Mattsson et al. [1] and supplementary material therein.

Sensitivity analysis 1

Figure S 1 shows the cost difference between the supergrid (S, S-Hland) scenarios and the continental (C, C-Hland) scenarios for a case with halved investment cost for transmission. The difference due to the supergrid option (i.e. the cost for C compared to S) is 5%, compared to 1% in the base scenario.

Figure S 2 shows the system cost when nuclear power is included as an investment option. The scenarios S-Nuc, C-Nuc and R-Nuc refer to Table 2 in the main text. The costs for the S-Nuc and C-Nuc scenarios are so similar that they cannot be distinguished in the figure. Compared to the base case (without the nuclear option), the regional scenario without transmission connections between subregions (R-Nuc) is feasible for the default land availability. In addition, it is a mere 10% more costly than the interconnected scenarios.

Figure S 3 shows the system cost when concentrated solar power (CSP) is included as an investment option, in addition to solar PV. The scenarios S-CSP, C-CSP and R-CSP, refer to Table 2 in the main text. The difference between the scenarios are similar to the base case (without CSP as a technology investment option). However, the system cost is slightly lower.

Note that the regional scenario, R-CSP is still infeasible for $CO_2$ targets below 10 g $CO_2$/kWh.

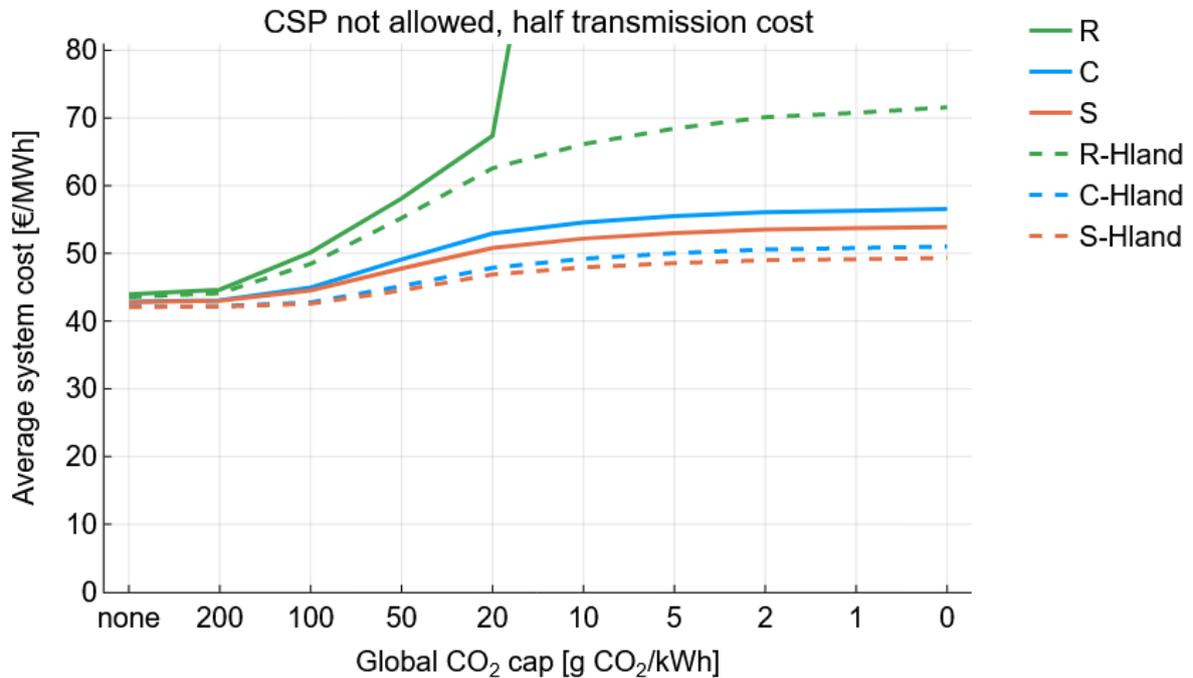

*Figure S 1 The system cost of electricity for increasingly stringent carbon targets, in the cases of Regional (R), Continental (C) and Superconnected (S) transmission expansion availability. Note that the investment costs for transmission (both converter station- and line costs) was halved compared to that in the base case. All cases were run for default land availability (R, C, S) and high land availability (R-Hland, C-Hland, S-Hland). For further details regarding the scenarios, see Table 2 in the main text.*

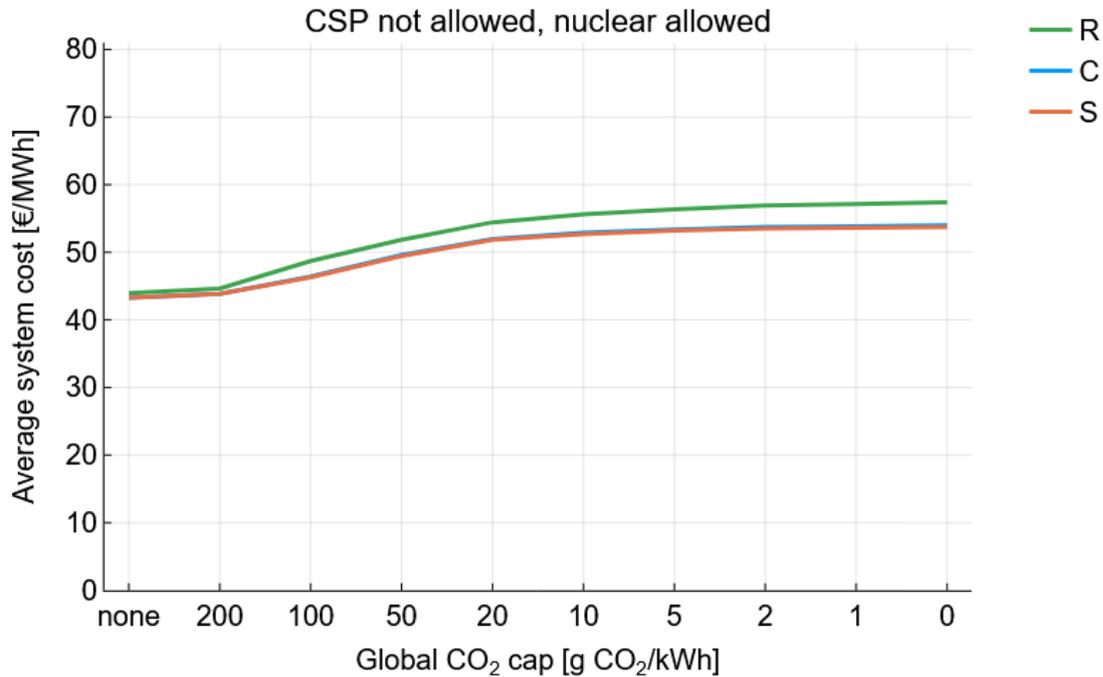

*Figure S 2 The system cost of electricity for increasingly stringent carbon targets, in the cases of Regional (R-Nuc), Continental (C-Nuc) and Superconnected (S-Nuc) transmission expansion availability. Note that nuclear is included as a technology option, which was not the case in the base scenario. All cases were run for default land availability. For further details regarding the scenarios, see Table 2 in the main text.*

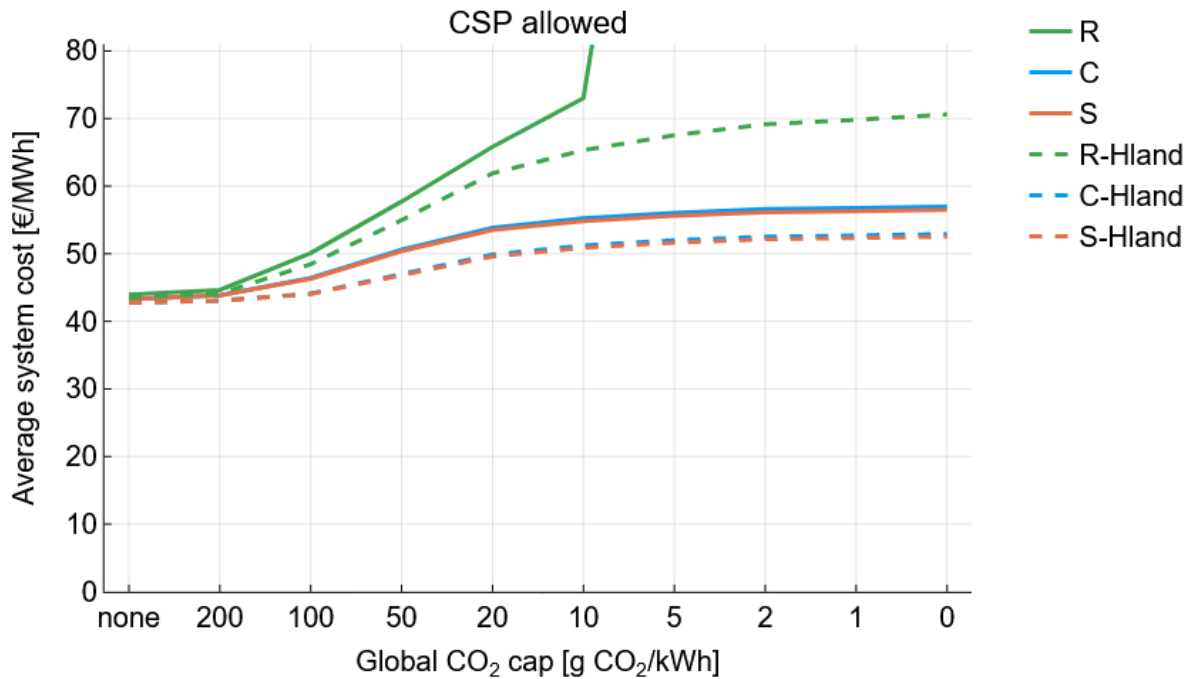

*Figure S 3 The system cost of electricity for increasingly stringent carbon targets, in the cases of Regional ('R-CSP'), Continental ('C-CSP') and Superconnected ('S-CSP') transmission expansion availability. Note that CSP is included as a technology option, which was not the case in the base scenario. All cases were run for default land availability. For further details regarding the scenarios, see Table 2 in the main text.*

Sensitivity analysis 2

Figure S 4 *The effect of using the SSP1, instead of the SSP2, scenario to construct the demand. The increase in total system cost for a system without interconnected superregions compared to with interconnected superregions, using varying assumptions for battery and solar PV costs. The left figure shows the difference between cases S and C, and the right side between cases S-Hland and C-Hland.* shows the effect of a higher demand for electricity. Instead of using the SSP2 scenario for constructing the electricity demand, the SSP1 scenario, with am electricity demand approximately 25% higher than that for SSP2 was used. See further details in the Method section of the main text. The effect of the higher demand is that the maximum cost reduction from the supergrid option (compared to continental grids) is increased from 4% to 5%. This occurs when the investment cost for both solar PV and storage are high.

Sensitivity analysis 3

Figure S 5 and Figure S 6 shows the cost reduction from the supergrid option (compared to continental grids) when using two different years of weather input data: 1998, which was a year with high wind output in Europe and Central Asia (Figure S 5) and 2003, which was a year with low wind output in Europe and Central Asia (Figure S 6).

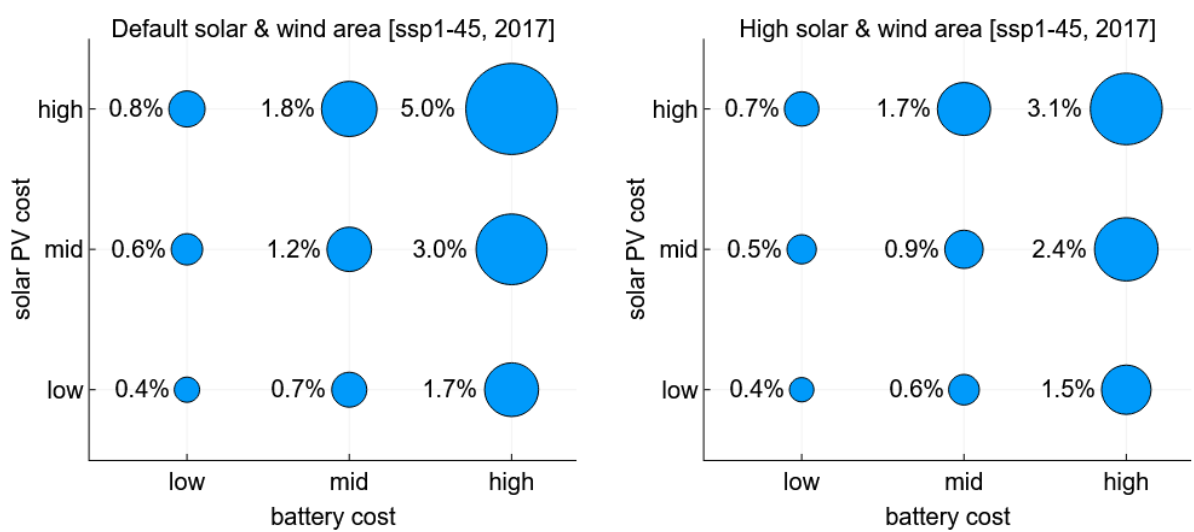

*Figure S 4 The effect of using the SSP1, instead of the SSP2, scenario to construct the demand. The increase in total system cost for a system without interconnected superregions compared to with interconnected superregions, using varying assumptions for battery and solar PV costs. The left figure shows the difference between cases S and C, and the right side between cases S-Hland and C-Hland.*

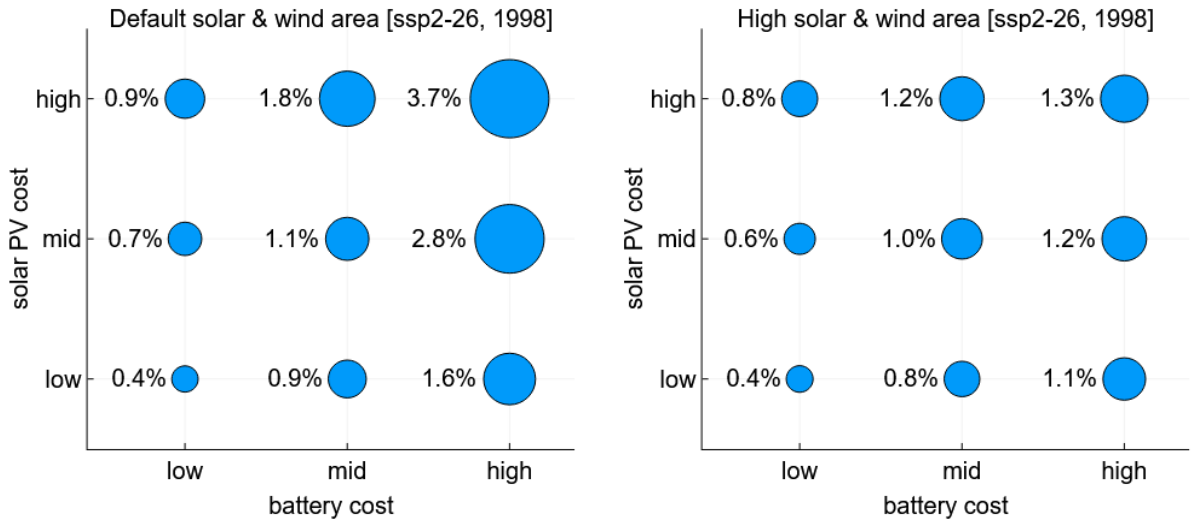

*Figure S 5 The effect of using the weather year 1998 as input data. The increase in total system cost for a system without interconnected superregions compared to with interconnected superregions, using varying assumptions for battery and solar PV costs. The left figure shows the difference between cases S and C, and the right side between cases S-Hland and C-Hland.*

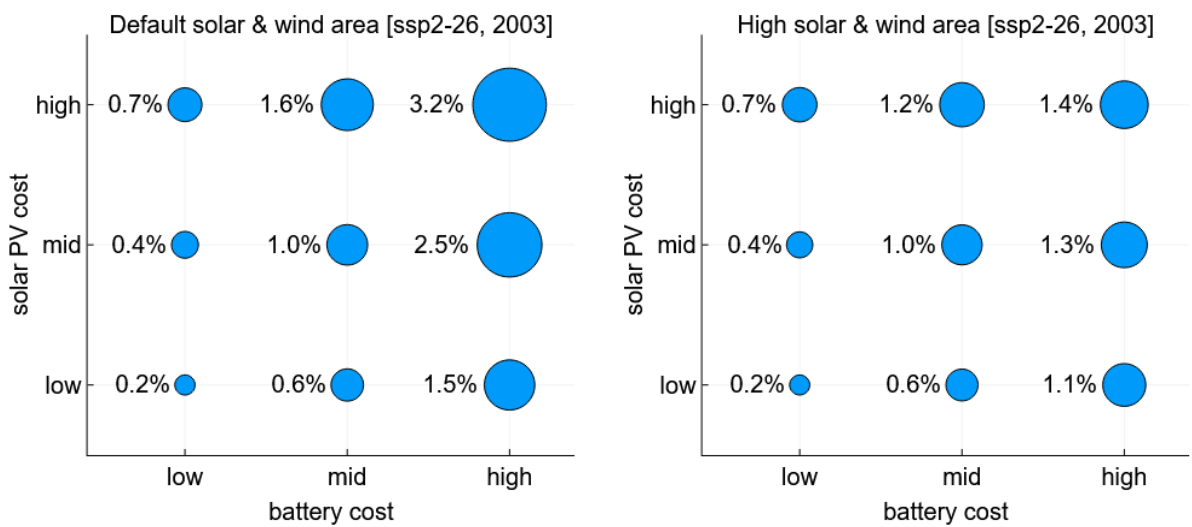

*Figure S 6 The effect of using the weather year 2003 as input data. The increase in total system cost for a system without interconnected superregions compared to with interconnected superregions, using varying assumptions for battery and solar PV costs. The left figure shows the difference between cases S and C, and the right side between cases S-Hland and C-Hland.*

Wind and solar resource characterization

Figure S 7 shows the wind and solar resource for the three superregions a) Europe, b) Central Asia and c) China. The resources are evaluated in terms of LCOE (y-axis) compared to the annual demand (x-axis). The figure shows that Central Asia has both wind- and solar resources that may deliver energy the equivalent of the total annual demand at an LCOE just slightly above 40 €/MWh. Europe has wind resources that can do the same at an LCOE of 60 €/MWh, and solar resources at 40 €/MWh. China, however, displays a scarcity of low-cost

wind resources, especially in the case of the default land availability (8% of remaining land available for wind farms), where only about 35% of demand may be satisfied using wind power at 60 €/MWh or less. For China, the less conservative land constraints ('Hland' in the scenarios in Table 2 in the main text), seem to *a priori*, have a large impact on the possibility to satisfy demand using wind power.

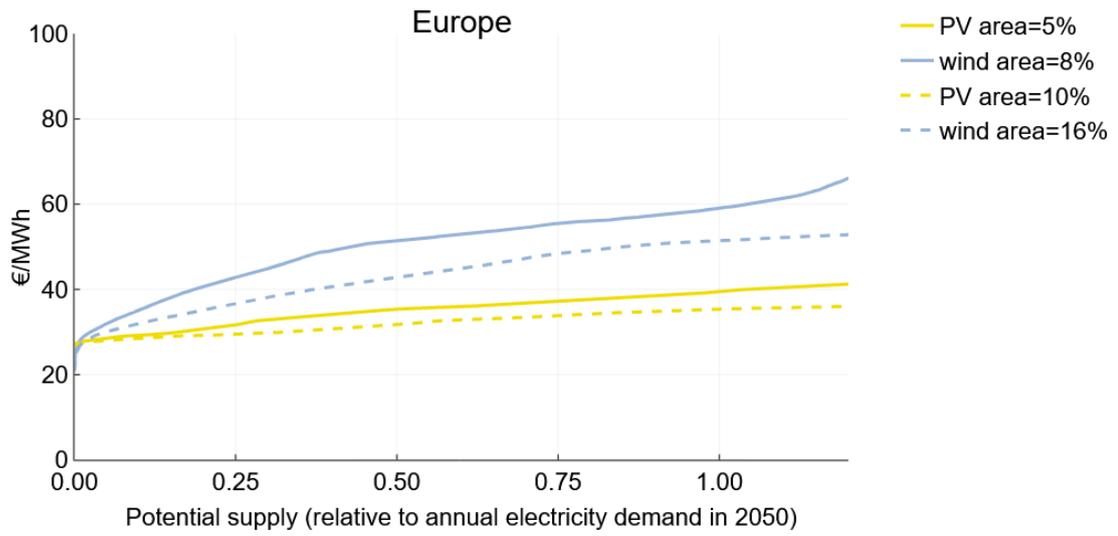

a)

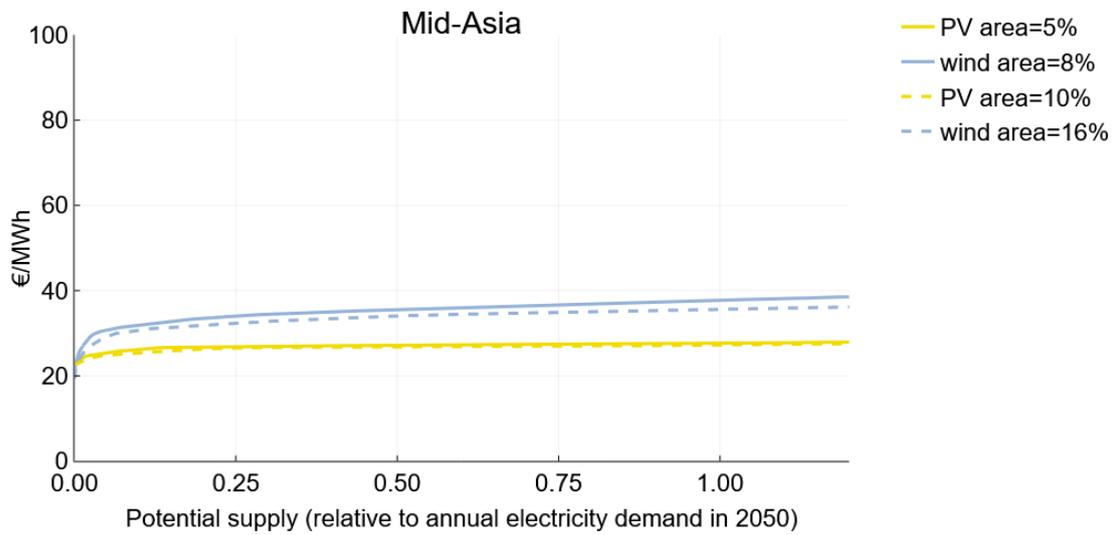

b)

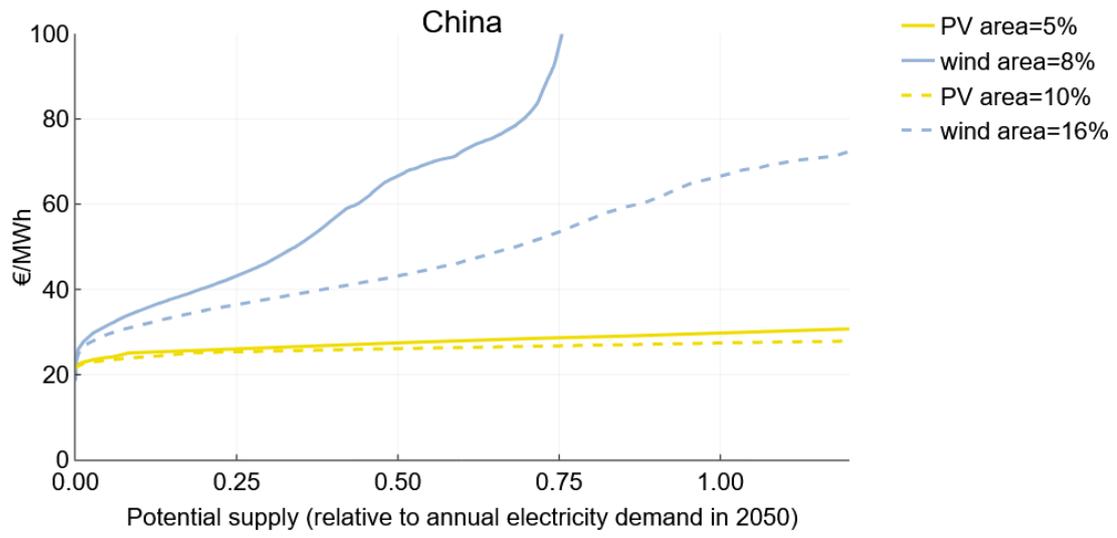

c)

*Figure S 7 Cost-supply curves of aggregate wind and solar resources in the three superregions a) Europe, b) Central Asia and c) China.*

Land exclusion masks and exploitable land

Figure S 8 shows land exclusion masks and remaining exploitable land for solar PV and CSP plants in three subregions a) FRA in Europe, b) KZH in Central Asia and c) CH_E in China. After the land masks have been applied, a fixed fraction of remaining land is considered exploitable for solar farms (5%/10% in default/high land availability scenarios).

Figure S 9 shows land exclusion masks and remaining exploitable land for onshore wind power in three regions a) FRA in Europe, b) KZH in Central Asia and c) CH_E in China. After the land masks have been applied, a fixed fraction of remaining land is considered exploitable for wind farms (8%/16% in default/high land availability scenarios).

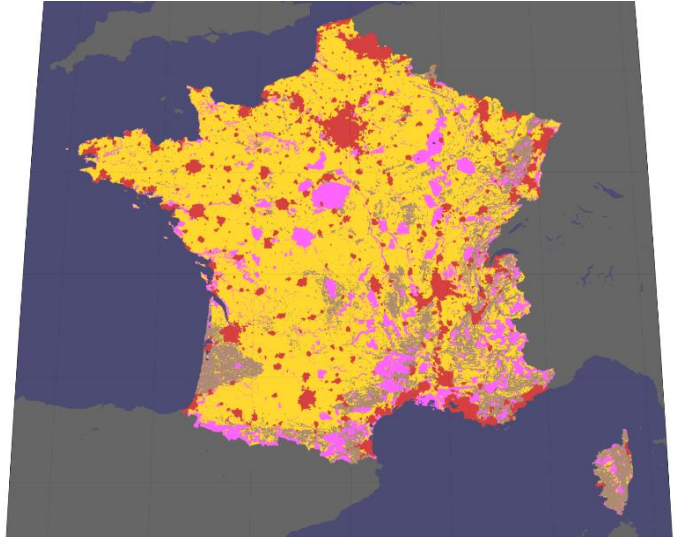

a)

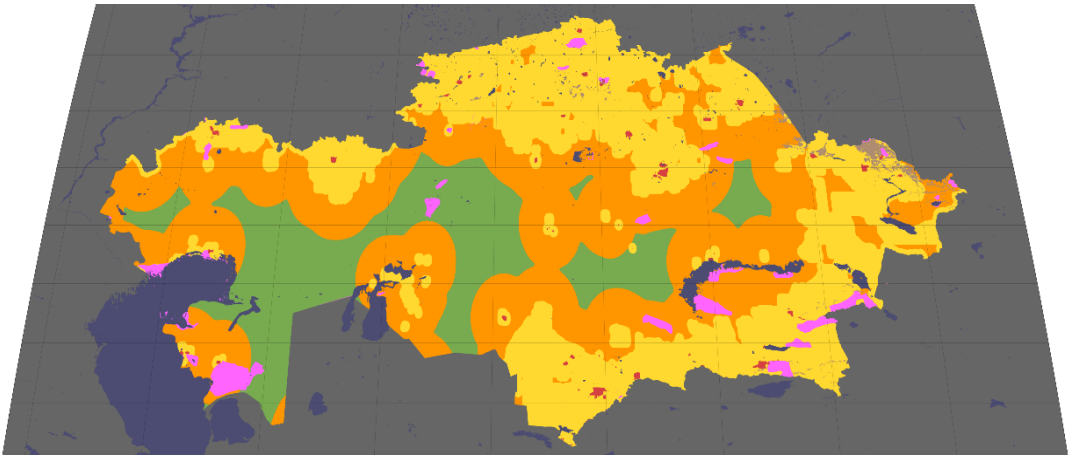

b)

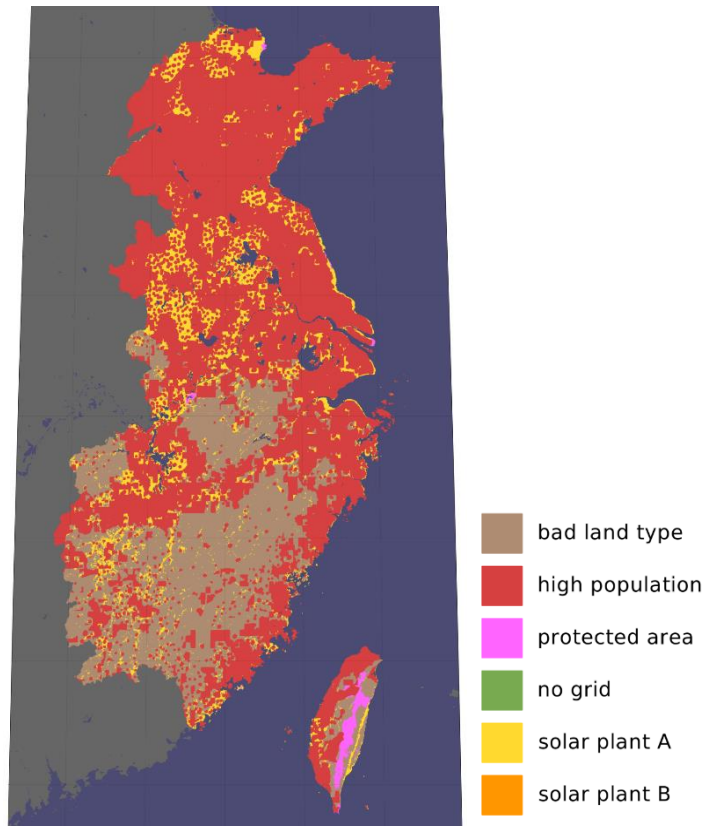

c)

*Figure S 8 Available land for solar PV and CSP plants after having excluded highly populated areas, forests, protected areas etc. (as described in the main text). Yellow colors denote remaining land area for solar power deployment, a fixed fraction of which is considered exploitable for solar farms (5%/10% in default/high land availability scenarios). Similarly, 5% of remaining land in highly populated areas (red) is allowed for solar rooftop PV.*

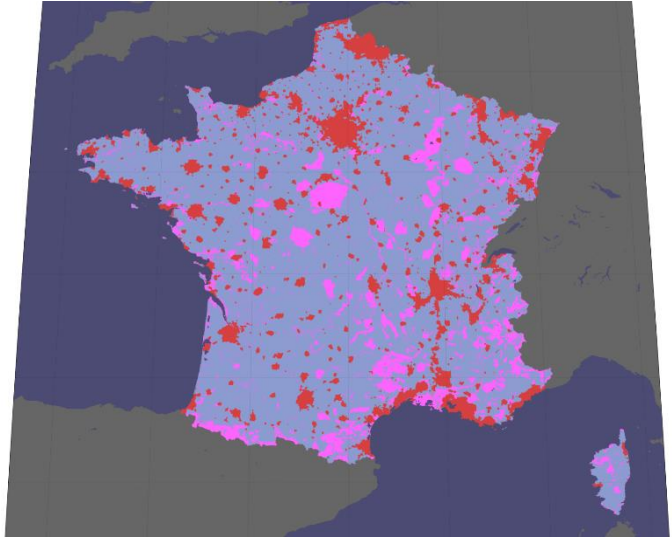

a)

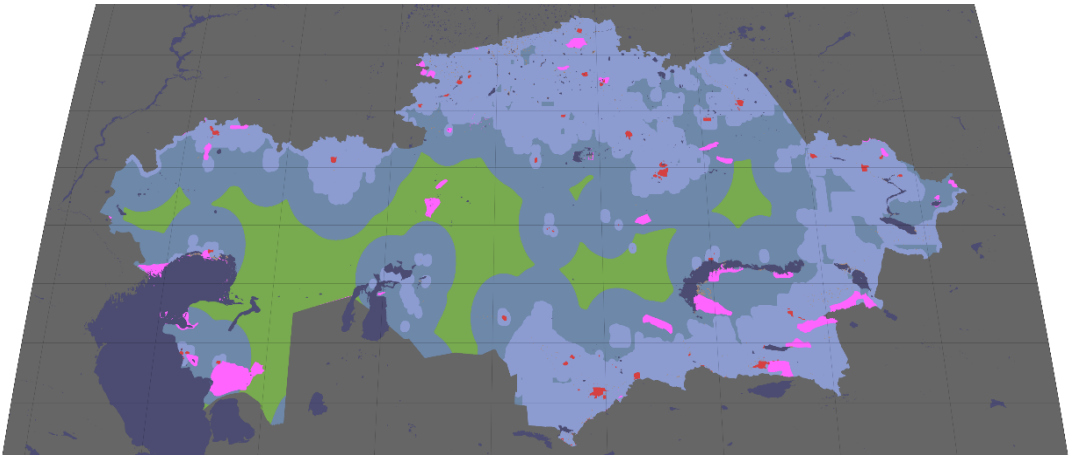

b)

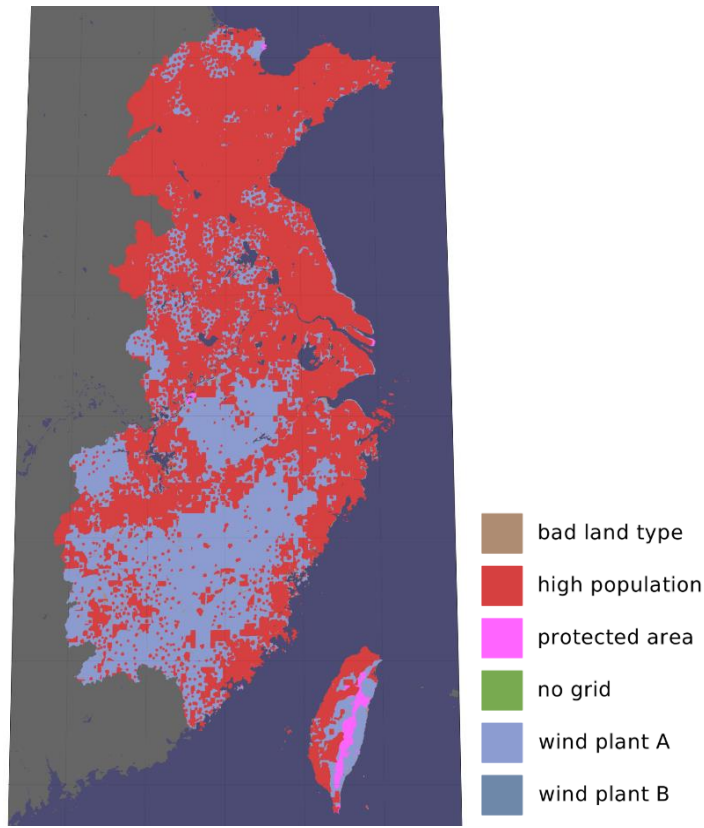

c)

*Figure S 9 Available land for wind power after having excluded highly populated areas, protected areas etc. (as described in the main text.) Blue colors denote remaining land area for wind power deployment, a fixed fraction of which is considered exploitable for wind farms (8%/16% in default/high land availability scenarios).*

Average regional wind speeds

Figure S 10 shows the average wind speeds for the three superregions. The average wind speeds were calculated by averaging over the 25% windiest sites each year, based on ERA5 data from ECMWF [ref 32 of main paper]. The year 2017 was used for the main results in the paper, while the years 2003 and 1998 were used for **Sensitivity analysis 3**. 2017 was selected because it represents a normal year, while 1998 was selected for its relatively high wind speeds in CAS and Europe and 2003 for its relatively low wind speeds in the same regions. As is apparent from Figure S 10, there is no obvious correlation between regions. In fact, it appears that years wind high wind speeds in Europe seem to display low wind speeds in China.

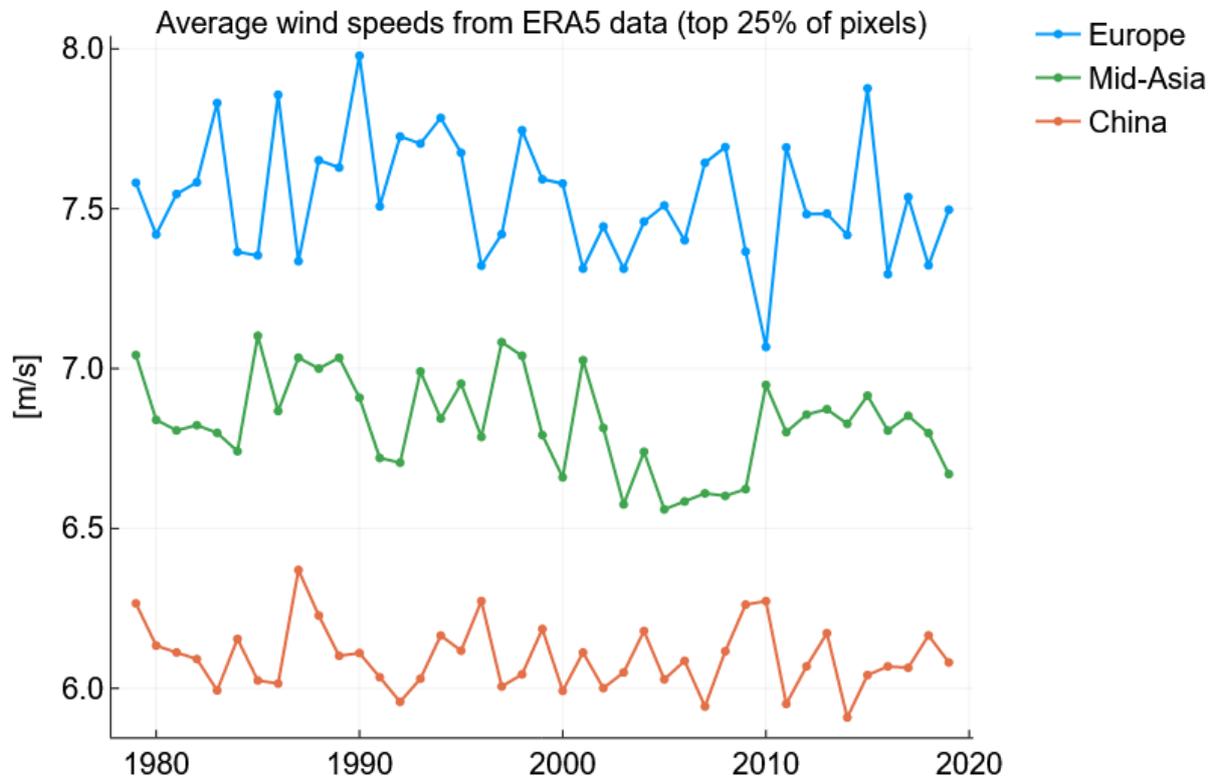

*Figure S 10 Average ERA5 wind speeds for the years 1979 to 2019 for the three superregions Europe, Mid-Asia and China, using the windiest 25% of pixels in each superregion.*

Exploited area for wind farms

One of the most critical parameter assumptions of the GIS procedure for estimating regional renewable potentials is the exploitable land area after areas for disallowed land types, protected areas and high population density have been removed. Below we show early results from a new study in which we use databases of real world wind turbine locations in several countries to evaluate this assumption. Our combined database covers all wind turbines in USA, Germany, Denmark and Sweden. Since wind penetration levels are quite high in Denmark (nearly 50% of national energy generation is based on wind power) and can locally reach similar levels in the other countries, our hope is to use this data to find a realistic upper level to exploitable area for large regions. In Figure S11 below, we show a histogram of the share of area exploited for wind power as a share of the total area of each municipality or county, after disallowed areas have been removed. We note that the exploited area rarely exceeds our global upper limit of 8% of the total area. Although there are some unusually small municipalities in which the actual exploited area significantly exceeds this level, the histogram suggests that our parameter assumption is adequate (or might be increased to around 10-12%).

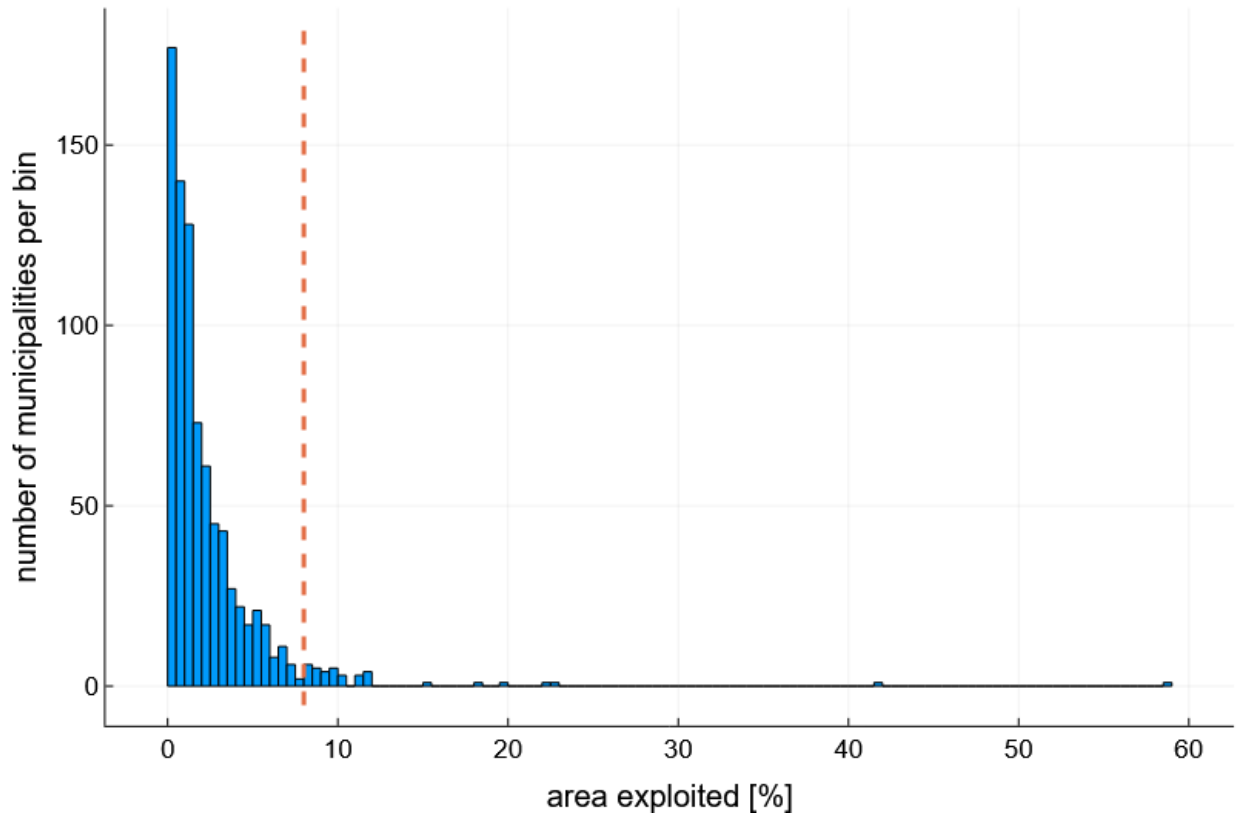

*Figure S 11 Distribution of exploited area for wind farms in all municipalities/counties of Germany, Denmark, Sweden and USA that have at least 10 MW of wind turbines (i.e. 838 of 3893 municipalities). The exploited area is calculated as the area currently dedicated to wind farms as a share of remaining area in each municipality after applying our GIS masks, i.e. after removing protected areas, wetlands and areas with too high population density. The dashed line represents our assumption that 8% of this remaining land area is exploitable for onshore wind farms.*